\documentstyle[12pt]{article}
\def\beqa{\begin{eqnarray}}
\def\eeqa{\end{eqnarray}}
\def\beq{\begin{equation}}
\def\eeq{\end{equation}}

\def\gd{g_{\mu\nu}}

\def\pa{\partial}

\def\alp{{\alpha}}

\def\gam{{\gamma}}

\def\ua{^{\alpha}}  
\def\ub{^{\beta}}

\def\umu{^{\mu}}
\def\unu{^{\nu}}

\def\dmunu{_{\mu\nu}}

\def\ie{{\it i.e. }}
\def\eg{{\it e.g. }}

\def\pr{{\it Phys. Rev.}\ }
\def\prl{{\it Phys. Rev. Lett.}\ }
\def\pl{{\it Phys. Lett.}\ }
\def\np{{\it Nucl. Phys.}\ }

\def\ijmp{{\it Int. Journ. Mod. Phys.}\ }

\def\cqg{{\it Class. Quantum Grav.}\ }

\def\apj{{\it Ap. J.}\ }

\def\nc{{\it Nuovo Cim.}\ }

\def\mnras{{\it Mon. Not. R. Ast. Soc.}\ }

\def\rmp{{\it Rev. Mod. Phys.}\ }

\def\ass{{\it Astr. and Space Sci.}\ }

\begin{document}
\def\bib#1{[{\ref{#1}}]}
\begin{titlepage}

\title{Cosmological perturbations in singularity--free, deflationary models} 

\author{{S. Capozziello$^a$, G. Lambiase$^a$, and G. Scarpetta$^{a,b}$} \\ 
{\em  $^a$Dipartimento di Scienze Fisiche ``E. R. Caianiello'',} \\
{\em Universit\`{a} di Salerno, I-84081 Baronissi, Salerno, Italy.} \\
{\em  $^a$Istituto Nazionale di Fisica Nucleare, Sezione di
Napoli,} \\
{\em $^b$International Institute for Advanced Scientific Studies,} \\
{\em Vietri sul Mare (SA) - Italy.}}  
	      \date{}
	      \maketitle
\begin{abstract}
We consider  scalar perturbations of energy--density for a class 
of cosmological models where an early phase of accelerated expansion 
evolves, without
any fine--tuning for graceful exit, towards the standard Friedman eras
of observed universe. The quantum geometric procedure 
which generates such models
agrees with results for string cosmology since it works if
dynamics is dominated by a primordial  fluid of extended massive objects.
The main result is that characteristic scales of cosmological interest, 
connected with the extension of such early objects, are  selected.
\end{abstract}
\maketitle
\vspace{20. mm}
e-mail address: \\
capozziello@vaxsa.csied.unisa.it \\
lambiase@vaxsa.csied.unisa.it \\
scarpetta@vaxsa.csied.unisa.it

\end{titlepage}

\section{\normalsize\bf Introduction}
Inflationary ``paradigm'' can be considered one of the main achievement
of recent cosmology since it solves a large amount of shortcomings
of standard cosmological model \bib{piran},\bib{brandenberger}. 
However, it is well known
that, among the several formulations of inflation, none is completely
satisfactory due to the fine tuning requests of each of them \bib{adams}.
Sometimes we have to avoid the extremely high rate of magnetic monopole
production \bib{guth}, sometimes we have to build a suitable 
scalar field potential in order to allow the slow rolling 
\bib{albrecht},\bib{linde};
in any case, we have the "graceful exit" problem since models continue to
inflate without recovering the standard today observed Friedman behaviour
\bib{piran},\bib{brandenberger}. Another great problem of many inflationary
models is that they are not singularity free (\eg power law inflation 
\bib{lucchin}) so the main shortcoming of standard model is not solved
at all. 

Despite of this state of art, inflation seems, up to now, the only
mechanism able to produce a perturbation spectrum that, starting from
initial quantum fluctuations, could reproduce the observed large scale 
structures of the universe \bib{mukhanov}.
However, in all inflationary models, the comparison of generated density 
perturbations with observational data strongly constrains the model
parameters. These limitations follows from the observed isotropy of cosmic 
microwave background radiation \bib{smoot}, in particular from the COBE
data \bib{cobe}. Most inflationary models predict that density perturbations 
are generated by the fluctuations of a scalar field (the inflaton) which are 
expanded to macroscopic sizes during the inflationary age. 

The further issue that any inflationary model has to satisfy is that, during 
the expansion, perturbations which are inside the Hubble radius $H^{-1}$
at the beginning of inflation expand past the Hubble radius and reenter it at
late times as large scale density perturbations.
To calculate the amplitude of density perturbations and to study the 
transition from inflationary to the Friedman era, it is necessary to know 
how the background geometry change with time.

Therefore, a coherent theory of early universe should:
\begin{enumerate}
\item be connected to some unification scheme of all interactions of nature;
\item avoid the initial singularity;
\item evolve smoothly, \ie without fine tuning, from an inflationary stage
to a decelerating Friedman era;
\item give rise to a perturbation spectrum in agreement both with the observed
microwave background isotropy and with the large scale structures.
\end{enumerate}
In other words, we search for a  a cosmological model,
 connected with some fundamental theory, that, at a certain epoch,
acquires a deflationary behaviour \bib{barrow} reproducing a suitable
perturbation spectrum.

In this paper, we face such a problem. By a quantum geometric procedure 
\bib{caianiello1}, we construct a class of cosmological models of 
deflationary type wich smoothly evolves towards Friedman epochs.
Over this background, we analyze the theory of gauge invariant cosmological 
perturbations for the density contrast 
${\displaystyle \frac{\delta \rho}{\rho}}$ connecting it with the scales of
astrophysical interest.

The main hypothesis to build such models is that the early universe is
dominated by a fluid of finite--size objects which give rise to a
dynamics very similar to that of string--dilaton cosmology 
\bib{veneziano},\bib{tseytlin}. However, the starting point is different from
that of string theory since our procedure is just a quantum 
geometric scheme.

Furthermore, we do not need any scalar field to implement inflation since the
proper size of extended objects and the geodesic embedding procedure from an 
eight--dimensional tangent fiber bundle ${\cal M}_{8}$ to the usual
${\cal V}_{4}$ manifold of general relativity naturally give rise to an 
exponential inflationary--like behaviour.

The paper is organized as follows.
In Sec.2, we describe the geometric procedure and the background model.
Sec.3 is devoted to the discussion of the deflationary behaviour through
the matter--energy density acting as the source in the Einstein equations.
In Sec.4, we construct the theory of gauge--invariant cosmological 
perturbations using the above models as background. The analysis is devoted 
to the large and small scale limits and then to the selection of scales
of astrophysical interest. Conclusions are drawn in Sec.5.

\section{\normalsize\bf Geodesic embedding and the background model}

The quantum geometric procedure and 
the background model which we are going to use
are treated in detail in \bib{caianiello1}. Here, we outline the main 
features which we need for cosmological perturbations. The starting point
is that if we consider dynamics of an extended massive object in general 
relativity, a limiting maximal acceleration, compatible with the size
$\lambda$ of the object and the causal structure of the spacetime manifold, 
emerges \bib{misner}.
Such a proper constant acceleration $A$ yields a Rindler horizon at a 
distance $|A|^{-1}$ from the extremity of the object in the longitudinal
direction. In other words, the parts of the object will be in causal contact
only if $|A|<\lambda^{-1}$. It is worthwhile to stress that the parameter
$A$ (or $\lambda$) is related to the "mass" of the extended object and
we are using physical units where $8\pi G=c=1$. 

Let us take into account a Friedman--Robertson--Walker (FRW) spacetime
whose scale factor, with respect to the cosmic time $t$, is $a(t)$.

By using the equation of geodesic deviation 
\bib{caianiello1},\bib{caianiello2},\bib{gasperini} we get that 
the size $\lambda$ of the object is compatible with the causal structure if
$|\lambda \ddot{a}/a|<1$.

Consequently, we have a maximal allowed curvature depending on $\lambda$
and the cosmological model becomes singularity free.
This fact is in sharp contrast with usual perfect fluid FRW cosmology where 
curvature and matter--energy density are singular in the limit $t\rightarrow 0$.
Then, the introduction of finite size objects, instead of pointlike 
particles, in primordial cosmological background modifies dynamics so that 
the singular structure of general relativity is easily regularized.
It is worthwhile to note that such a feature does not depend on the 
particular background geometry which we are considering.
More formally, a causal structure in which proper accelerations cannot exceed
a given value $\lambda^{-1}$ can be imposed over a generic spacetime
${\cal V}_{4}$ regarding such a manifold as a four--dimensional
hypersurface locally embedded in a eight--dimensional tangent fiber bundle
${\cal M}_{8}$, with metric
\beq
g_{AB}=\gd\otimes\gd\,,
\eeq
and coordinates $x^{A}=(x\umu,\lambda u\umu)$, where 
${\displaystyle u\umu=\frac{dx\umu}{ds}}$ is the usual four velocity
and $\mu,\nu=1,...,4,\;\;\;\;A,B=1,...,8$ 
\bib{caianiello1},\bib{caianiello2}.

The embedding of ${\cal V}_{4}$ into ${\cal M}_{8}$, determined by the eight
parametric equations $x\umu=x\umu(\xi\ua)$ and $u\umu=u\umu(\xi\ua)$,
gives rise to a spacetime metric $\tilde{g}\dmunu(\xi)$, locally induced by 
the ${\cal M}_{8}$ invariant interval
\beq
\label{2.1}
d\tilde{s}^2=g_{AB}dx^Adx^B=
\gd\left(dx\umu dx\unu+\lambda^2du\umu du\unu\right)
\equiv\tilde{g}\dmunu d\xi\umu d\xi\unu\,,
\eeq
where
\beq
\label{2.2}
\tilde{g}\dmunu=g_{\alp\beta}
\left(\frac{\pa x\ua}{\pa\xi\umu}\frac{\pa x\ub}{\pa\xi\unu}+
\lambda^2\frac{\pa u\ua}{\pa\xi\umu}\frac{\pa u\ub}{\pa\xi\unu}\right)\,.
\eeq

Let us now take into consideration a FRW background modified by such a 
geodesic embedding\footnote{This geometric procedure is called "geodesic
embedding" since the velocity field $u\umu(\xi)$, solution of the geodesic 
equations, defines the embedding of ${\cal V}_{4}$ into ${\cal M}_{8}$.}.

In conformal coordinates $\xi\umu=(\eta,\vec{x})$, a FRW flat metric is
\beq
\label{2.3}
\gd(\xi)=\mbox{diag}[a^2(\eta)(1,-1,-1,-1)]\,,
\eeq
where ${\displaystyle d\eta=\frac{dt}{a}}$ defines the conformal time. The
velocity field for an extended object comoving in this background is
\beq
\label{2.4}
u\umu(\xi)=\left(a^{-1},0,0,0\right)\,,
\eeq
By Eqs.(\ref{2.1}) and (\ref{2.2}), the geodesic embedding gives
rise to the new metric
\beq
\label{2.5}
\tilde{g}\dmunu(\xi)=
\mbox{diag} a^2\left(1+\lambda^2\frac{{a'}^2}{a^4},-1,-1,-1\right)\,,
\eeq
corrected by a $\lambda^{2}$ term with respect to (\ref{2.3}). The prime 
indicates the derivative with respect to $\eta$.
The cosmic time results now
\beq
\label{2.6}
t=\int d\eta\left(a^2+\lambda^2\frac{{a'}^{2}}{a^2}\right)^{1/2}\,,
\eeq
or, in terms of the scale factor only,
\beq
\label{2.7}
t=\lambda\int\frac{da}{a}\left(1+\frac{a^4}{\lambda^2{a'}^{2}}\right)^{1/2}\,.
\eeq
The Hubble parameter is now
\beq
\label{2.8}
H=\frac{\dot{a}}{a}=\left(\frac{a'}{a}\right)
\left[a^2+\lambda^2\left(\frac{a'}{a}\right)^2\right]^{-1/2}\,,
\eeq
with the limiting value
\beq
\label{2.9}
H_{0}=\lambda^{-1}\,,
\eeq
for ${\displaystyle 
\left(\lambda^{2}\left(\frac{a'}{a}\right)^2\gg a^2\right)}$.

It is easy to see that the scale factor, with respect to the cosmic time
$t$, has an initial exponential growth
which regularly evolves towards a standard Friedman behaviour.

\section{\normalsize\bf Deflationary behaviour of energy--density}
Modified geometry implies an initial de Sitter behaviour which is not
connected with dynamics of some scalar  field but it simply comes from
the presence of extended (and massive) objects. The $e$-folding number,
\ie the duration of inflation, and the horizon scale depend on the size 
$\lambda$ without any initial value problem or fine tuning.

The natural scale to which to compare perturbations is $\lambda$:
they are inside the Hubble radius if they are smaller than $\lambda$
while they are outside it if they are greater than $\lambda$.
In other words, $\lambda$ determines the crossing time
(either out of the Hubble radius or into the Hubble radius).

Considering the $(0,0)$--Einstein equation for a spatially flat model,
we have 
\beq
\label{3.1}
H^2=\frac{\rho}{3}\,,
\eeq
so that
\beq
\label{3.2}
\rho=3\left(\frac{a'}{a}\right)^2
\left[a^2+\lambda^2\left(\frac{a'}{a}\right)^2\right]^{-1}\,.
\eeq
Immediately we see that
\beq
\label{3.3}
\rho\simeq\frac{3}{\lambda^2}\,,\;\;\;\;\;\;\mbox{for}\;\;\;\;
\lambda^2\left(\frac{a'}{a}\right)^{2}\gg a^2\,,
\eeq
and
\beq
\label{3.4}
\rho\simeq 3\frac{{a'}^{2}}{a^4}\,,\;\;\;\;\;\;\mbox{for}\;\;\;\;
\lambda^2\left(\frac{a'}{a}\right)^2\ll a^2\,.
\eeq
The first case corresponds to an effective cosmological constant
${\displaystyle \Lambda=\frac{3}{\lambda^2}}$ selected by the mass 
(\ie the size) of the primordial extended objects \bib{caianiello1}; 
the second case
is recovered as soon as 
the universe undergoes the
 post--inflationary reenter phase. We stress again the fact that such
a behaviour does not depend on the specific form of the scale factor $a$
 and the deflationary phase
is smooth.

As in \bib{caianiello1}, we can couple dynamics with ordinary fluid matter in
order to obtain a more realistic cosmological scenario. In doing so, 
we have to consider a perfect fluid state equation
\beq
\label{3.5}
p=(\gamma-1)\rho\,,
\eeq
which, using also the contracted Bianchi identity in FRW spacetime gives
the continuity equation
\beq
\label{3.6}
\dot{\rho}+3H(p+\rho)=0\,.
\eeq
By Eqs(\ref{3.5}) and (\ref{3.6}), we get
\beq
\label{3.7}
\rho=Da^{-3\gamma}\,.
\eeq
For the sake of simplicity, $\gamma$ is assumed constant. It defines
the thermodynamical state of the fluid and it is related to the sound speed
being $\gam-1=c_{s}^{2}$. 
By inserting this fluid into the Einstein equations, 
the scale factor of the universe, expressed in conformal time is
\bib{barrow},\bib{caianiello1},\bib{turok}
\beq
\label{3.8}
a(\eta)=a_{0}\eta^{2/(3\gamma-2)}\,,
\eeq
where
$a_{0}$ is a constant depending on $\lambda$ and $\gamma$.

The matter--energy density results, from Eq.(\ref{3.2}),
\beq
\label{3.10}
\rho=3\left(\frac{2}{3\gamma -2}\right)^2\frac{1}{\eta^2}
 \left[a_{0}^2\eta^{\frac{4}{3\gamma-2}}+
\lambda^{2}\left(\frac{2}{3\gamma-2}\right)^2\frac{1}{\eta^2}\right]^{-1}\,,
\eeq
from which $\rho\sim$ constant for 
${\displaystyle \frac{\lambda}{\eta^{6\gamma/(3\gamma -2)}}\gg 1}$ and 
${\displaystyle \rho\sim\eta^{\frac{6\gamma}{2-3\gamma}}}$ in the opposite
case. 
The standard situations for $\gamma=4/3$ (radiation dominated regime)
and $\gamma=1$ (matter dominated regime) are easily recovered.
It is interesting to see that it is not only the specific value of 
$\gamma=0$, as usual, that allows to recover inflation but, mainly,
the scale $\lambda$. In the regime 
${\displaystyle \frac{\lambda}{\eta}\gg 1}$, the constant matter density
value is independent of $\gamma$.

In the next section, we shall study the density contrast
${\displaystyle \frac{\delta\rho}{\rho}}$ which gives the perturbation 
spectrum. Due to the smooth transition from the inflationary to the FRW
regime, the perturbation scale lengths do not need any cut--off and can be
parametrized in all their evolution by the parameter $\lambda$
which has to be compared with Hubble causal horizon $H^{-1}$.

\section{\normalsize\bf Gauge--invariant cosmological perturbations}

In the gauge-invariant formalism, the conformal invariance 
and the frame-independence are requested for variables connected to 
perturbations in order to eliminate the pure gauge modes \bib{mukhanov}.
Furthermore, any generalized theory of gravity can be recast
into the Einstein theory plus one or more than one additional scalar 
fields \bib{hwang}. In some sense, our quantum geometric procedure
can be seen as a modified theory of gravity.

We can turn now to consider the  scalar perturbations.
For a spatially flat FRW metric, the line element is \bib{mukhanov}
\beq
ds^2=a^2(\eta)\left[(1+2\phi)d\eta^2-2B_{;i}dx^{i}d\eta-
dx^{i}dx^{j}(2E_{;ij}+(1-2\psi)\delta_{ij})\right]\,.
\eeq
It is always possible to construct combinations of the scalar quantities
$\phi,\psi,E,B$ which are invariant under general coordinate transformations
as $x\ua\rightarrow\tilde{x}\ua+\xi\ua$. A useful combination,
which gives rise to the  invariant perturbation potentials, is
\beqa
\Phi&=&\phi+\frac{1}{a}[(B-E')a']'\\
\Psi&=&\psi-\frac{a'}{a}(B-E')\,.\nonumber\\
\eeqa
Such a choice simplifies the evolution equations for density perturbations
and, as we shall see below, furnishes quantities with a clear physical 
meaning.
In the same way, we can construct perturbed Einstein equations which are
invariant under general coordinate transformations and, consequently,
we get gauge invariant quantities. These equations are generally  written in 
terms of $\Phi$ and $\Psi$. Furthermore, the symmetries of the stress--energy
tensor can give additional simplifications. In fact, as it is clearly shown 
in \bib{mukhanov}, if the source stress--energy tensor  is symmetric,
we have $\Phi=\Psi$, so that we need 
just one evolution equation (plus, however, the gauge choice).

Usually, $\Phi$ is called the "gauge-invariant potential" and characterizes
the amplitude of scalar density perturbations. It is a function of the 
conformal time $\eta$ and the spatial coordinates ${\bf x}$. It is important
to note that below the Hubble radius $H^{-1}$, $\Phi$ has the role of a
Newtonian potential for the density contrast yielded by perturbations.

The general gauge--invariant evolution equation for 
scalar adiabatic perturbations
is \bib{mukhanov},\bib{bardeen}
\beq
\label{master}
\Phi''+3{\cal H}(1+c_{s}^{2})\Phi'
-c_{s}^{2}\nabla^{2}\Phi+
\left[2{\cal H}'+(1+3c_{s}^{2}){\cal H}^{2}\right]\Phi=0\,,
\eeq
${\cal H}$ is the Hubble 
parameter in the conformal time defined as
\beq
{\cal H}=\frac{a'}{a}\,.
\eeq
$c_{s}$, as above,  is the sound speed.
It is interesting to note that Eq.(\ref{master}) can be recast in terms of
the scale factor $a$ by the variable change $dt=a\,d\eta$ and $da/dt=aH$
\bib{caldwell}. In this way, the information contained in the evolution
equation is directly related to the background. However, for our purposes,
it is better to use the "conformal time picture" 
since it immediately shows when the  
sizes of perturbations are comparable to the characteristic scale 
length $\lambda$.

Another important step is the decomposition of the perturbation potential
into spatial Fourier harmonics
\beq
\label{fourier}
\Phi(\eta,{\bf x})=\int d^3 k
\tilde{\Phi}(\eta,{\bf k})e^{i{\bf k}\cdot{\bf x}}\,,
\eeq
where $k$ is the wavenumber.
Essentially, this decomposition consists 
in replacing $\nabla^{2}\rightarrow -k^{2}$ in the dynamical equation
(\ref{master}). It allows to follow the evolution of a single mode. In
our case,  considering a specific mode, we can follow it from the
inflationary deSitter stage to the deflationary Friedman era.
For example, if $k\ll H$, we have long wavelength modes which furnish the 
spectrum of perturbations during inflation. In our case, it is interesting
to compare such modes with the "natural" scale of the model, that is
$H_{0}=\lambda^{-1}$.

Before performing the Fourier analysis, it is useful to simplify the
dynamical problem
by a suitable change of variables. Eq.(\ref{master}) 
can be reduced to the simpler form
\beq
\label{master1}
u'' - c^2_s\nabla^2 u - {\theta''\over\theta}u=0\,,
\eeq
where $\theta$ is
\beq
\theta={1\over a}\left(\frac{\rho_0}{\rho_0 + p_0}
\right)^{1/2}={1\over a}\left(\frac{1}{1 +  
p_0 /\rho_{0}}\right)^{1/2}=\frac{1}{a\sqrt{\gamma}}\,,
\eeq
and the gauge--invariant gravitational potential $\Phi$ is given  by 
\beq
\Phi= {1\over2}(\rho_0+ p_0)^{1/2}u\,.
\eeq
From now on, the subscript $"_0"$ will indicate the unperturbed quantities.

The density perturbations are given by 
\beq
\label{contrast}
\frac{\delta\rho}{\rho_0}=
\frac{2\left[\nabla^2\Phi-3{\cal H}\Phi'-
3{\cal H}^2\Phi\right]}{3{\cal H}^2}\,.
\eeq
In the specific case  we are considering, using the solution
(\ref{3.8}), we get
\beq
\label{theta}
\theta(\eta)=\left[\frac{2}{H_{0}\gamma^{1/2}(3\gamma-2)}\right]
\eta^{2/(3\gamma-2)}\,,
\eeq
and
\beq
\frac{\theta''}{\theta}=\left[\frac{6\gamma}{(3\gamma-2)^{2}}\right]
\frac{1}{\eta^2}\,.
\eeq
After the Fourier trasform, Eq.(\ref{master1}) becomes
\beq
\label{master2}
u''_{k}+
\left[c_{s}^2 k^2-\frac{6\gamma}{(3\gamma-2)^{2}\eta^2}\right]u_{k}=0\,,
\eeq
which is nothing else but a Bessel equation.
The density perturbations can be rewritten as
\beq
\label{contrast1}
\frac{\delta\rho}{\rho_{0}}=-(\rho_{0}+p_{0})^{1/2}
\left[\left(1+\frac{k^2}{{\cal H}^2}\right)u_{k}(\eta)+
\frac{u'_{k}(\eta)}{\cal H}\right]\,,
\eeq 
where
\beq
\label{hubble}
{\cal H}=\left(\frac{2}{3\gamma-2}\right)\frac{1}{\eta}\,.
\eeq

The general solution of (\ref{master2}) is
\beq
\label{solution}
u_{k}(\eta)=\eta^{1/2}\left[A_{0}J_{\nu}(z)+B_{0}Y_{\nu}(z)\right]\,,
\eeq
where $J_{\nu}(z)$ and $Y_{\nu}(z)$ are Bessel functions and
\beq
\nu=\pm\frac{3\gamma+2}{2(3\gamma-2)}\,,\;\;\;\;\;\;\;
z=c_{s}k\eta\,.
\eeq
$A_{0}$ and $B_{0}$ are arbitrary constants.

\vspace{2. mm}

Actually, we are interested in the asymptotic behaviour of $\Phi$,
that is $u_{k}$, since it, by (\ref{contrast}), determines the large scale 
structures of the universe. 

The large scale limit is recovered as soon as $k^2\ll 
\theta''/\theta$, or $k\ll H$.
This means that the solution (\ref{solution}) becomes
\beq
\label{lsl}
u_{k}(\eta)\simeq \eta^{1/2}\left[\frac{A_{0}}{\Gamma(\nu+1)}
\left(\frac{c_{s}k\eta}{2}\right)^{\nu}-
\frac{B_{0}\Gamma(\nu)}{\pi}\left(\frac{c_{s}k\eta}{2}\right)^{-\nu}
\right]\,.
\eeq
For different values of $\gamma$, the index $\nu$ can be positive or negative
determininig growing or decaying modes. 

In the vacuum--dominated era $(\gamma=0)$, we have, for $k\rightarrow 0$,
\beq
u_{k}(\eta)\sim\left[ B_{0}\sqrt{\frac{2}{\pi c_{s}}}\right]k^{-1/2}\,,
\eeq
or
\beq
\frac{\delta\rho}{\rho_{0}}\sim 
\left[B_{0}\sqrt{\frac{2(\rho_{0}+p_{0})}{\pi c_{s}}}\right]k^{-1/2}\,.
\eeq
This is a nice feature since the spectrum of perturbations is a constant
with respect to $\eta$ as it must be during inflation, when dynamics
is frozen \bib{kolb}. As we pointed out, we recover the case $\gamma=0$
any time that $H_{0}=\lambda^{-1}=k_{\lambda}$, that is the feature of the 
spectrum is fixed by the natural scale of the model\footnote{ To be more 
precise, by using (\ref{contrast1}), we get
$$ \frac{\delta \rho}{\rho_{0}}\sim\sqrt{\rho_{0}+p_{0}}
\left(1+\frac{k^{2}}{3H^2}\right)u_{k}(\eta)\,.$$ 
As soon as $k^{2}\ll H^2$,
in particular $k^{2}\ll k_{\lambda}^{2}$, the long wavelength
perturbations go beyond the horizon and their dynamics results frozen.
This feature is always present during inflation. In our case, it is
recovered without any fine--tuning.}. 

\vspace{2. mm}

If $\gamma$ is any, in particular $\gamma=1,4/3,2$, 
corresponding to the 
cases "dust", "radiation" 
and "stiff matter" 
respectively, we get
\beq
\label{matter}
u_{k}(\eta)\sim \frac{A_{0}}{\Gamma(\nu+1)}\left(\frac{c_{s}}{2}\right)^{\nu}
\eta^{(\nu+1/2)}k^{\nu}-\frac{B_{0}\Gamma(\nu)}{\pi}
\left(\frac{c_{s}}{2}\right)^{-\nu}\eta^{(1/2-\nu)}k^{-\nu}\,.
\eeq
In particular, for $k\rightarrow 0$, only the second term survives.
The density contrast, in the same limit, is
\beq
\label{contrast3}
\frac{\delta \rho}{\rho_{0}}\sim\frac{B_{0}\Gamma(\nu)}{\pi}
\sqrt{\rho_{0}+p_{0}}\left(\frac{2}{c_{s}}\right)^{\nu}
\frac{(3\gamma-2)^2}{12}\eta^{(\frac{5}{2}-\nu)}k^{(2-\nu)}\,.
\eeq
It is interesting to note that, for $\gamma=1$, $\nu=5/2$ and
\beq
\frac{\delta\rho}{\rho_{0}}\propto k^{-1/2}\,,
\eeq
that is we lose the time dependence also if the scales are reentered the
horizon (for $\gamma=1$ we are in the Friedman regime).

\vspace{2. mm}

The small scale limit is recovered as soon as in (\ref{master1}) or
(\ref{master2})
$k^{2}\gg \theta''/\theta$. The solution can be written as
\beq
u_{k}(\eta)\sim\sqrt{\frac{2}{\pi c_{s}k}}\left[A_{0}\cos(c_{s}k\eta)+
B_{0}\sin(c_{s}k\eta)\right]\,,
\eeq
and  looking at (\ref{contrast1}),  also the density contrast is an 
 oscillating function in $\eta$.
From a cosmological point of view, this limit is not very interesting 
since it is not directly connected to dynamics of inflation.

\section{\normalsize\bf Discussion and conclusions}

In this paper, we have constructed the theory of gauge--invariant
cosmological perturbations for a  model in which, by a
geometric procedure of local embedding, the metric is modified. 

Such a modification can be read
as the effect of a fluid of extended primordial objects whose dynamics
alters the cosmological background. 
A very important point is that the size of the objects gives rise to
an inflationary period that smoothly evolves toward a Friedman era.
 
Also the cosmological perturbations are affected by such a dynamics since
the scales (\ie the wavenumbers $k$) are regulated by the size $\lambda$
which is a natural scale giving the Hubble horizon $H_{0}=\lambda^{-1}$
during inflation.
Then the limits to compare {\it very} large scale structures and
{\it small} large scale structures are $k\ll H_{0}$ and $k\gg H_{0}$.
In other words, $\lambda$ fixes the time at which perturbations
cross the horizon and reenter, enlarged, into it without any
fine--tuning. This point has to be discussed in detail comparing it
with the standard method used to calculate the amplification of
perturbations after reenter.  

In the limit $k\ll H$ and for adiabatic perturbations, the quantity
\beq
\label{conservation}
\zeta=\frac{2}{3\gamma}\left(\Phi+ H^{-1}\dot{\Phi}\right)+\Phi\,,
\eeq
or its Fourier transform
\beq
\label{conservation1}
\tilde{\zeta}=\frac{2}{3\gamma}
\left(\tilde{\Phi}+ {\cal H}^{-1}\tilde{\Phi}'\right)
+\tilde{\Phi}\,,
\eeq
is conserved \bib{mukhanov},\bib{caldwell}.

In such a limit, Eq.(\ref{master}) corresponds to $\dot{\zeta}=0$,
so that, for long wavelengths, the use of $\zeta$ to obtain the evolution
of $\Phi$ is justified. However, this position holds only on scales
larger than Hubble radius (when $c_{s}^2\nabla^2\Phi$ is negligible)
and not for all dynamics. At very early and very late times, it is realistic
to neglect also the derivative $\dot{\Phi}$ \bib{capozziello}, so that
we have
\beq
\label{pegaso}
\Phi(t_{f})=\left[\frac{1+\frac{2}{3}\gamma^{-1}(t_{f})}{
1+\frac{2}{3}\gamma^{-1}(t_{i})}\right]\Phi(t_{i})\,,
\eeq
which means that
the amplitudes of perturbations crossing out from the Hubble radius
and reentering it later are related. The net change is due to the state 
equation $p=(\gamma-1)\rho$ describing the model before crossing and after 
reenter. 
As $\gamma\rightarrow 0$, the amplification becomes huge solving the problem 
that microscopic perturbations enlarge to macroscopic (astronomical) sizes.

In any case, 
$t_{i}$ must be taken well before the beginning of inflation and well
after its end.  Then, if Eq.(\ref{pegaso}) is a useful tool to calculate
how inflation enlarges the amplitude of primordial perturbations, it
gives rise to a further fine--tuning problem since $t_{i}$ and $t_{f}$,
and the relative $\gamma(t_{i})$ and $\gamma(t_{f})$, must be chosen
with a lot of care. 

Our model bypasses such a shortcoming since inflation smoothly comes to an
end and also the amplitude of perturbations smoothly evolves towards
the Friedman era. However, due to the presence of extended objects at very
beginning, the model starts as inflationary and singularity--free  
so that, from a cosmological point of view,
we do not have an epoch before inflation.
Besides, the size $\lambda$ triggers the scales at which galaxies and
cluster of galaxies should form \bib{kolb}. 

The quantum--geometric procedure  which we used acquires physical 
meaning only if
we suppose that, in an early phase, the contribution of finite--size
objects becomes dominant. 

In a forthcoming paper, we shall discuss
some physically motivated examples of such dynamics.

\newpage

\begin{center}
{\bf References}
\end{center}
\begin{enumerate}
\item\label{piran}
D.S. Goldwirth and T. Piran, {\it Phys. Rep.} {\bf 214} (1992) 223. 
\item\label{brandenberger}
R.H. Brandenberger, \rmp {\bf 57} (1985) 1.
\item\label{adams} 
F.C. Adams, K. Freese, and A.H. Guth, \pr {D 43} (1991) 965
\item\label{guth}
A.H. Guth, \pr {\bf D 23} (1981) 347.
\item\label{albrecht}
A. Albrecht and P. Steinhardt, \prl {\bf 48} (1982) 1220.
\item\label{linde}
A.D. Linde, \pl {\bf B 108} (1982) 380.\\
A.D. Linde, \pl {\bf B 120} (1983) 177.
\item\label{lucchin}
F. Lucchin and S. Matarrese, \pr {\bf D 32} (1985) 1316.
\item \label{mukhanov}
V.F. Mukhanov, {\it Sov. Phys.} JETP {\bf 68}, 1297, 1988.\\
V.F. Mukhanov, H.A. Feldman, R.H. Brandenberger, {\it Phys. Rep.} {\bf 215},
203, 1992.
\item\label{smoot}
G.F. Smoot {\it et al.}, \apj {\it Lett.} {\bf 396}, L1 (1992).\\
K. Ganga {\it et al.}, \apj {\it Lett.} {\bf 410}, L57 (1993). 
\item\label{cobe}
R. Scaramella, N. Vittorio, \apj {\bf 353}, 372, 1990.\\
J.R. Gott, C. Park, R. Juszkiewicz, W.E. Bies, D.P. Bennet, F.R.
Bouchet, A. Stebbins, \apj {\bf 352}, 1, 1990.\\
{\it The Cosmic Microwave Background: 25 Years Later},
ed. N. Mandolesi, N. Vittorio, Kluwer, Dordrecht, 1990.\\
R. Scaramella, N. Vittorio, \apj {\bf 375}, 439, 1991.
\item\label{barrow}
J.D. Barrow \np {B 310} (1988) 743.\\
J.E. Lidsey, \pr {\bf D 55} (1996) 3303.
\item\label{caianiello1}
E.R. Caianiello, M. Gasperini, and G. Scarpetta, \cqg
{\bf 8} (1991) 659. \\
E.R. Caianiello,  A Feoli, M. Gasperini, and G. Scarpetta, {\it Int. J. Theor.
Phys. } {\bf 29} (1990) 131. \\
E.R. Caianiello, M. Gasperini, and G. Scarpetta, {\it Il Nuovo Cimento} {105B}
(1990) 259.
\item\label{caianiello2}
E.R. Caianiello, {\it Lett. Nuovo Cimento} {\bf 32} (1981) 65.\\
E.R. Caianiello, {\it La Rivista del Nuovo Cimento} {\bf 15} n.4 (1992).
\item\label{gasperini}
M. Gasperini, \ass {\bf 138} (1987) 387.
\item\label{veneziano}
G. Veneziano, \pl {\bf B 265} (1991) 287.\\
M. Gasperini, J. Maharana, and G. Veneziano, \pl {\bf B 272} (1991) 277.\\
K.A. Meissner and G. Veneziano, \pl {\bf B 267} (1991) 33.
\item\label{tseytlin}
A.A. Tseytlin, \ijmp {\bf A 4} (1989) 1257.\\
A.A. Tseytlin and C. Vafa, \np {\bf B 372} (1992) 443.
\item\label{misner}
C.W. Misner, K.S. Thorne, and J.A. Wheeler {\it Gravitation}
ed. Freeman, S. Francisco (1973).
\item\label{turok}
N.Turok, \prl {\bf 60}, 549 (1988).
\item\label{hwang}
J. Hwang, \cqg {\bf 7}, 1613, 1990.\\
J. Hwang, \pr {\bf D 42}, 2601,1990.\\
G.F.R. Ellis, M. Bruni, \pr {\bf D 40}, 1804, 1989.\\
G. Chibisov, V. Mukhanov, \mnras, {\bf 200}, 535, 1982.
\item\label{bardeen}
J. Bardeen, \pr {\bf D 22} (1980) 1882.
\item\label{caldwell}
R.R. Caldwell, \cqg {\bf 13} (1996) 2437.
\item\label{kolb}
E.W. Kolb, and M.S. Turner {\it The Early Universe} \\
Addison-Wesley 1990 (Redwood City, Calif.)\\
P.J.E. Peebles {\it Principle of Physical Cosmology},\\
Princeton Univ. Press 1993 (Princeton).
\item\label{capozziello}
S. Capozziello, M. Demianski, R. de Ritis, and C. Rubano
\pr {\bf D 52} (1995) 3288.\\
S. Capozziello and R. de Ritis, \nc {\bf B 109} (1994) 783.

\end{enumerate}
\vfill
\end{document}